\documentclass[reprint, prper,amsmath,amssymb,aps]{revtex4-2}

\usepackage{url}
\usepackage{graphicx}
\usepackage{dcolumn}
\usepackage{bm}
\usepackage{hyperref}
\usepackage[mathlines]{lineno}
\usepackage[english]{babel}

\begin{document}

\preprint{Preprint submitted to \textit{Physical Review Physics Education Research}}

\title{Using Large Language Models to Analyze Engagement in Computational Thinking via Computational Physics Essays}

\author{Sean Savage}
\affiliation{Department of Physics and Astronomy, Purdue University, West Lafayette, IN, 47907, U.S.A.}

\author{Amir Bralin}
\affiliation{Department of Physics and Astronomy, Purdue University, West Lafayette, IN, 47907, U.S.A.}
\affiliation{Department of Physics and Astronomy, Texas Tech University, Lubbock, TX, 79409, U.S.A.}

\author{Paul Hur}
\affiliation{Department of Physics, Freie Universit\"at Berlin, Berlin, Germany.}

\author{N. Sanjay Rebello}
\affiliation{Department of Physics and Astronomy, Purdue University, West Lafayette, IN, 47907, U.S.A.}
\affiliation{Department of Curriculum and Instruction, Purdue University, West Lafayette, IN, 47907, U.S.A.}

\begin{abstract}
As computational thinking (CT) becomes increasingly important to physics education, the need for authentic, project-based assessments has grown. While open-ended multimodal assignments, such as Computational Physics Essays (CPEs), help capture student reasoning and encourage active learning, they introduce a significant evaluation bottleneck. Manually grading these complex notebooks across a complex taxonomy of computational practices is resource-intensive and limits scalability in large-enrollment courses. In this study, we investigated the viability of using a multimodal Large Language Model (LLM) to automate the evaluation of 100 student-generated CPEs. Using a human-coded baseline, we systematically evaluated the model's capacity to detect student engagement across 20 distinct CT sub-practices and a holistic overall quality score. The results showed that the LLM performs very well on clearly defined tasks, achieving an 84\% exact agreement with human raters on the binary sub-practices. However, more subjective constructs proved challenging, with the model reaching only a 71\% agreement for the holistic quality analysis. Our findings demonstrated that while LLMs can reliably automate the detection of specific computational practices, subjective evaluation remains a hurdle.
\end{abstract}

\maketitle

\section{\label{sec:intro}Introduction}

Computation has become a fundamental domain of science, recognized by the American Association of Physics Teachers (AAPT) as the ``third pillar" of physics, alongside experiment and theory \citep{AAPT2016}. This paradigm shift has elevated the importance of fostering Computational Thinking (CT) in STEM education. First popularized by Jeanette Wing in 2006 as the thought processes involved in ``solving problems, designing systems and understanding human behavior, by drawing on the concepts fundamental to computer science, the definition of CT has since expanded to encompass distinct modeling practices specific to mathematics and science \citep{wing2006computational, shute2017demystifying, weintrop2016computationalpractices}. Computation is rapidly moving outside the classroom as computational tools become more available in the workplace, creating an essential need to ensure students can apply these tools correctly \citep{wang2021integrating}. This priority is reflected not only by AAPT but also in broader STEM standards, such as the Next Generation Science Standards (NGSS) and the Accreditation Board for Engineering and Technology (ABET) \citep{NGSS2013, abet2019}.

Despite the demand for more CT across physics education, a significant gap remains in how CT is operationally measured in higher education. As a result of this gap, the integration of computation in undergraduate physics often suffers from students defaulting to unproductive problem-solving habits. In their work on mathematical reasoning in physics, Tuminaro and Redish \citep{reddish2007} identified the ``plug-and-chug" epistemic frame. In this mindset, students treat problem-solving as a localized game: identify a target variable, search for a corresponding formula, and insert numbers. When computation is introduced without careful pedagogical design, students naturally transfer this mathematical frame to the computer. They treat programming as just a high-powered calculator rather than a scientific sensemaking tool \citep{aiken2013i}. Consequently, students generate correct numerical outputs, but they do so without any of the authentic computational skills they will need in professional practice \citep{Kortemeyer2016-sb}. In the context of generative artificial intelligence (AI), we define `authentic computational thinking' not merely as the successful execution of syntax, but as the metacognitive process of physical sensemaking. It requires students to explicitly justify their mathematical assumptions, evaluate their model's limitations, and actively connect their code back to the physical constraints of the system.

The increasing availability and adoption of generative AI has severely exacerbated this challenge. Large language models (LLMs) can now generate functional Python scripts instantly, allowing students to completely bypass the cognitive effort needed for algorithmic design \citep{li2023structuredchainofthoughtpromptingcode, zion2024, becker2023, wang2024examining}. This capability renders traditional coding assignments potentially ineffective in assessing genuine student understanding. Therefore, to verify learning in this age of AI, assessments must demand \textit{deep context}, where computation is linked to a specific physical scenario, and \textit{personalized modeling}, which requires students to make, evaluate, and iteratively defend design decisions.

To address this challenge, educators have turned to complex, project-based assignments such as the Computational Physics Essay (CPE). Adapted from the framework established by Odden and Caballero \citep{odden2019}, the CPE is a multimodal document blending code, data visualization, and narrative text into a scientific argument within a Jupyter Notebook. 

However, evaluating these open-ended, multi-modal assignments presents a scalability bottleneck. Historically, physics education research has relied heavily on multiple-choice instruments to measure conceptual understanding at scale \citep{hestenes1992force, thornton1998assessing, singh2016multiple}. While multiple-choice assessments excel at identifying broad trends, they are fundamentally limited in capturing students' mental models and reasoning processes \citep{Kuechler_Simkin_2003, roediger2005}. Written explanations and computational assignments provide a much richer window into student cognition \citep{nieswandt2009written, mcneill2011supporting}; however, evaluating them is notoriously resource-intensive. Achieving acceptable inter-rater reliability across highly granular computational sub-practices requires many hours of qualitative human coding, forcing researchers and instructors in large-enrollment courses to choose between the scalability of superficial assessments and the richness of qualitative data \citep{kortemeyer2023, casalino2021, BUTCHER2010489}. 

Recent advancements in LLMs offer a promising solution to this bottleneck. Previous research has demonstrated the feasibility of using AI to evaluate short problem solutions or identify narrow student misconceptions \citep{kortemeyer2025, nlp2024misconceptions, Wan_Chen_2024, ssavage2025}. However, testing the capacity of an LLM to reliably parse a multimodal essay, where code, narrative prose, and visualizations overlap across a wide array of sub-practices, remains largely unexplored. 

This study addresses this gap by shifting the traditional analytical focus toward automated rubric-based evaluation. Rather than utilizing an LLM to generate code, we deploy a custom-prompted, multimodal LLM pipeline to analyze a pre-existing human ground-truth dataset of 100 multimodal Computational Physics Essays. We systematically evaluate the model's capacity to detect engagement across 20 granular computational sub-practices, as well as its alignment with a holistic ``Overall CT Quality" score used by expert human raters to capture deep physical sensemaking. Specifically, we address the following research questions:

\begin{enumerate}
    \item \textbf{RQ1:} How reliably can an LLM identify distinct sub-practices of computational thinking within multimodal student essays compared to expert human raters?
    \item \textbf{RQ2:} To what extent can the LLM capture more subjective nuances of assessment, such as the overall holistic CT quality score, compared to explicit structural markers?
\end{enumerate}
\section{Background \& Framework}\label{sec:background}

\subsection{Constructionism and Project-Based Learning}

Our pedagogical motivations regarding the hybrid-modality design of CPEs (computation + narrative rhetoric) are grounded in constructionism. This framework shows that deep conceptual mastery is not transmitted from instructor to student, but actively built on by the learner via public artifacts \citep{papert1991constructionism}. In this context, we define the `artifact' as the CPE itself, rather than just the output/visual simulation from the code. This integrated essay of code, narrative text, and data visualization is contained in a Jupyter Notebook. 
We further base our decision to use CPE on the theory of project-based learning (PBL). As described by Krajcik and Shin \citep{krajcik2014pbl}, PBL environments promote scientific reasoning by centering learning around students' extended period of inquiry. The CPE serves as a capstone project for the end of the semester; Unlike weekly laboratory and recitation exercises, where the procedure is prescriptive. This extended duration allows students to engage in iterative improvement and gives them the opportunity to use the notebook as a canvas to test their hypotheses. 
Notably, this construction occurs in collaborative groups of 2-3 students, extending the pedagogical framework to social constructivism \citep{vygotsky1978mind}. While we hold that authentic Computational Thinking involves the social negotiation required to generate code \citep{obsniuk2015}, we explicitly acknowledge a boundary condition of our assessment methodology: the LLM pipeline evaluates only the static, final Jupyter notebook artifact. Therefore, this study does not measure the social negotiation process, but rather the codified evidence of physical sensemaking that results from it.

\subsection{The Computational Essay (CE)}
To operationalize our approach of constructionism and project-based learning, we utilize the computational essay framework established by Odden and Caballero \citep{odden2019}. A Computational Essay (CE) is a multimodal digital document that blends executable code, data visualizations, and narrative text into a cohesive scientific argument. Authored within interactive environments such as Jupyter Notebooks, CEs allow students to present their code not as a hidden ``black box,'' but as a transparent, integral component of their physical reasoning. Prior research has shown that this format successfully scaffolds professional scientific practices and fosters epistemic agency, as students must explicitly justify their computational choices within the text rather than simply producing a final numerical output \citep{Odden2023-tb}.

\subsection{Adapting the CE for Introductory Physics}
While the original computational essay framework was highly successful for upper-level undergraduate STEM majors \citep{odden2019}, deploying it in a large-enrollment, introductory mechanics context required specific pedagogical adaptations. To adapt the CE for this introductory population, we simplified the complexity of the computational methods students were expected to use. 

Our primary constraint was requiring the conceptual synthesis of at least two fundamental physics principles (e.g., conservation of energy and conservation of momentum) to model a system of interest. This constraint shifted the assignment's focus from broad computational creativity to targeted systems integration, ensuring that students solved physical problems using computation as a tool rather than focusing exclusively on advanced software engineering. We term this adapted assessment the Computational Physics Essay (CPE). This pedagogical adaptation ensured that the resulting CPEs provided measurable evidence of physics-specific computational skills appropriate for first-year students.

\subsection{Assessment Framework}

To measure evidence of students' engagement in CT in our essays, we required a consistent and specific definition of computational thinking. While numerous frameworks exist, most notably the comprehensive review by Shute et al. \citep{shute2017demystifying},  we found these generalist models insufficient for our investigation. Shute's framework is fundamentally rooted in traditional computer science, and therefore, it breaks down CT into general categories like decomposition and debugging. Although it performs well at evaluating how well a student constructs software, it lacks the necessary granularity to evaluate how well a student models real-world physics. As such, we adopt the taxonomy developed by Weintrop et al. \citep{weintrop2016definingcomputationalthinking}, which is explicitly designed for using computational methods in mathematics and science classrooms. Unlike other frameworks \citep{brennan2012new,shute2017demystifying}, this model elevates Data practices and Modeling \& Simulation Practices to primary domains. This helps us distinguish ``Computer Science" CT from more ``physics" CT (correct real-world modeling over syntax).

However, as noted by Weller \citep{weller2022}, the broad universality of Weintrop's taxonomy, which includes 22 distinct sub-practices, requires adaptation based on the context in order to be properly measured. Hence, we redefined his framework by identifying 20 specific sub-practices visible in the student's final Jupyter notebook (CPE). We explicitly excluded \textit{Programming} and \textit{Troubleshooting and Debugging}, from our list, both sub-practices of the \textit{Computational Problem Solving} Domain. These were removed as \textit{programming} is inherently a part of what the CPE artifact is, and \textit{Debugging} would generally not be seen in their final product, though it should be noted to be an important part of the process. The 20 specific sub-practices and the operationalized criteria used to score them can be seen in Table \ref{tab:rubric_full}.

\subsubsection{Data Practices}
In the context of undergraduate physics, data practices involve both the initial collection of real-world data and the handling of information generated via algorithms and physics equations. Because students were not collecting data experimentally via hardware, they researched theoretical conditions (e.g., drag coefficients) from external sources. Notably, students who solely cited an LLM (e.g., ChatGPT) as their data source did not receive credit for \textit{Collecting Data} in our analysis, as this bypasses authentic information gathering. Additionally, students generated arrays of data based on the information they collected and manipulated the generated data to create visualizations, such as plots, and used them to create meaningful conclusions about the problem they were trying to solve.

\subsubsection{Modeling \& Simulation Practices}
Where real-world experiments are constrained by hardware, modeling and simulation allow students to explore complex scenarios, such as adding non-linear air resistance to a projectile. To evaluate students' proficiency in this domain, we analyzed how they constructed and refined their computational models. An example of this involves writing iterative update loops, such as the Euler-Cromer method (i.e., a key technique for dealing with differential equations), to predict the future state of the system based on its current state and physical principles. Iterative loops are much more powerful than merely plugging numbers into an algebraic formula, allowing the model to handle varying forces and behave more realistically, and thus, it can represent a higher-level, more advanced modeling approach.

\subsubsection{Computational Problem Solving Practices}
Here, students focus on developing the programming skills needed to convert physics concepts into functional code. This domain measures a student's ability to decompose complex physical problems into logical steps. Key practices involve developing modular solutions and creating computational abstractions. In practice, this included instances like creating reusable Python functions for each of the forces used in their equations. For example, a student effectively utilizing abstractions defined variables like mass (m) and change in time (dt) at the start of their notebook rather than hard-coding numerical values into their equations, making the entire model easily adaptable.

\subsubsection{Systems Thinking Practices}
Physical systems often become difficult to understand when isolating properties, making a systems-level perspective essential. For these practices, we look for evidence of students defining clear system boundaries, such as having the Earth included when calculating the gravitational potential energy of a falling object. This domain assesses how well students synthesize micro-level interactions, like friction, into macro-level behavior, such as the total energy dissipated due to said friction. Furthermore, these practices connect to a core principle of science: how the information is communicated. Measuring how effectively students explain these intricate system interactions to the reader of the notebook is crucial, as it further proves they are actively thinking about how to make their ideas clear.

\subsection{Automated Evaluation via Large Language Models}

Historically, automated evaluation in physics education research has been confined to highly structured, objective formats, such as multiple-choice conceptual inventories or strictly formatted numerical inputs \citep{singh2016multiple}. While early natural language processing (NLP) tools expanded this capability by evaluating short-answer constructed responses, these traditional algorithms fundamentally relied on keyword matching and struggled to evaluate open-ended, multi-step scientific arguments \citep{lee_liu_2019}. The evaluation of a Computational Physics Essay, which interleaves executable Python code, explanatory narrative prose, and generated data visualizations into a single canvas, presents a significantly higher barrier. Traditional NLP cannot bridge the semantic gap between a block of code and the accompanying visual graph it generates. However, recent breakthroughs in Vision-Language Models (VLMs) and multimodal Large Language Models have demonstrated the ability to process interleaved text and image inputs simultaneously \citep{jiang2025}. By utilizing these multimodal capabilities, this study investigates whether an LLM can move beyond superficial keyword matching to holistically evaluate the complex, hybrid artifacts that characterize modern computational physics.

Evaluating open-ended, complex written responses at scale is notoriously resource-intensive \citep{kortemeyer2023}. Achieving acceptable inter-rater reliability often requires many hours of qualitative coding, forcing researchers and instructors to choose between the scalability of superficial assessments and the richness of qualitative data \citep{casalino2021, BUTCHER2010489}.

To evaluate complex written responses at scale, advancements in generative AI offer new options. Recent research has demonstrated the feasibility of using AI to evaluate problem solutions \citep{kortemeyer2023, kortemeyer2025}, identify student misconceptions \citep{nlp2024misconceptions}, and assist in generating feedback \citep{Wan_Chen_2024, latif2023fine}. Building upon this work, this study investigates the viability of LLMs as a scalable tool for evaluating open-ended CT assessments.

Rather than utilizing LLMs to simply evaluate code correctness, we deployed a custom-prompted LLM \citep{Chen2023} to categorize how participants frame and reason through computational physics problems. Because CPEs rely heavily on visual context, such as embedded Python code snippets and generated system diagrams, we specifically utilized a multimodal model for visual context. The model is guided using a structured prompt that incorporates role-playing, instructing the model to act as an expert physics education researcher, and structured reasoning prompts. By applying the same rubric criteria used by human raters directly to the LLM, we propose that generative AI can serve as a scalable mechanism for formative assessment, enabling instructors to efficiently track the integration of computational and physical reasoning.

\begin{table*}[htbp]
\caption{The Computational Physics Essay (CPE) Rubric. The rubric consists of 20 sub-practices adapted from Weintrop et al. (2016). Each item is scored Binary (0/1).}\label{tab:rubric_full}
\renewcommand{\arraystretch}{1.6} 
\scriptsize
\begin{ruledtabular}
\begin{tabular}{l l l} 
\textbf{Sub-Practice} & \textbf{Description (Criteria for Score = 1)} & \textbf{Student Example} \\
\colrule

\textbf{I. Data Practices} & & \\
1. Collecting Data & 
\parbox[t]{7.2cm}{Student gathers or records data from a real-world experiment, published dataset, or online source. \textit{Note: Citations of LLMs (e.g., ChatGPT) do NOT count. Links to external data DO count.}} & 
\parbox[t]{5.5cm}{``The following values for v (0-10 m/s) and mu (0.1-0.6) come from our resources below, such as Union Pacific."} \\

2. Creating Data & 
\parbox[t]{7.2cm}{Student generates new data points through computation, simulation, or transformation. \textit{Note: Examples include time steps, numerical arrays, and parameter sweeps. Setting a single variable to a value does NOT count.}} & 
\parbox[t]{5.5cm}{\texttt{t\_df = [0.01, 0.25...]; t\_s = [t * 86400 for t in t\_d]}} \\

3. Manipulating Data & 
\parbox[t]{7.2cm}{Student performs operations on data sets such as filtering, converting units, or reorganizing data. \textit{Note: Basic mathematical conversions, array reformatting, or unit changes applied across a dataset count as manipulation.}} & 
\parbox[t]{5.5cm}{ \texttt{if forces[i] < fatal\_limit:} \newline \texttt{\hspace*{2mm}plt.scatter(...)}} \\

4. Analyzing Data & 
\parbox[t]{7.2cm}{Student identifies trends, compares quantities, or interprets patterns in data. \textit{Note: Requires a mathematical calculation or deep physical interpretation. Vague surface-level observations do NOT count.}} & 
\parbox[t]{5.5cm}{``While this graph shows the crunch zone needed... major injuries can still happen."} \\

5. Visualizing Data & 
\parbox[t]{7.2cm}{Student uses graphs, plots, or tables to display or summarize computational results.} & 
\parbox[t]{5.5cm}{\texttt{plt.plot(t\_d, masses, linestyle='--')}} \\
\colrule

\textbf{II. Modeling Practices} & & \\
6. Using Models to Understand & 
\parbox[t]{7.2cm}{Student employs computational or mathematical models to represent or explain a physical phenomenon. \textit{Note: Code must loop or simulate physics. Simple calculator math does NOT count.}} & 
\parbox[t]{5.5cm}{\texttt{max\_res = f\_d - force\_drag; print(max\_res)}} \\

7. Using Models to Test & 
\parbox[t]{7.2cm}{Student uses the model to make predictions, test hypotheses, or explore outcomes.} & 
\parbox[t]{5.5cm}{Testing relationships: \newline \texttt{while h\_max <= 3.1: \# multiple iterations}} \\

8. Assessing Models & 
\parbox[t]{7.2cm}{Student reflects on the accuracy, validity, or limitations of model results, including physical assumptions.} & 
\parbox[t]{5.5cm}{``While the force calculation was more advanced, the model was incomplete... ignoring Moon curvature."} \\

9. Designing Models & 
\parbox[t]{7.2cm}{Student defines the model's components, interactions, and assumptions/equations before coding.} & 
\parbox[t]{5.5cm}{``The major constraint is to have the machine stop within the defined time parameter."} \\

10. Constructing Models & 
\parbox[t]{7.2cm}{Student writes, modifies, or describes code/rules to create a new model or extend an existing one.} & 
\parbox[t]{5.5cm}{\texttt{energy = 0.5 * m\_head * v\_i ** 2}} \\
\colrule

\textbf{III. Problem Solving} & & \\
11. Preparing Problems & 
\parbox[t]{7.2cm}{Student explicitly describes their strategy for breaking down the problem or outlines steps to write code. \textit{Note: Must explicitly outline a forward-looking strategy, plan, or structural breakdown.}} & 
\parbox[t]{5.5cm}{``In this first iteration, we estimated... In the fourth, we included gravitational potential energy."} \\

12. Choosing Tools & 
\parbox[t]{7.2cm}{The student justifies the selection of libraries, coding methods, or physics approaches. \textit{Note: Must include a functional description or convenience statement. Merely listing a tool without description does NOT count.}} & 
\parbox[t]{5.5cm}{``A point-particle system is most effective since translational speed is all that is relevant."} \\

13. Assessing Approaches & 
\parbox[t]{7.2cm}{Student compares different solutions, algorithms, or iterations and justifies why one is better. \textit{Note: Merely stating a syntax error was fixed does NOT count. Basic comparisons of code logic are accepted.}} & 
\parbox[t]{5.5cm}{``In iteration one we were unable to determine force... For iteration two we decided to use impulse."} \\

14. Modular Solutions & 
\parbox[t]{7.2cm}{Student structures code into reusable or iterative components, such as using for/while loops. \textit{Note: Linear scripts or sequential math steps do NOT count. Must include a reusable custom function or an explicit loop.}} & 
\parbox[t]{5.5cm}{\texttt{def calculate\_pe(mass, height):} \newline \texttt{\hspace*{2mm}return mass * g * height}} \\

15. Creating Abstractions & 
\parbox[t]{7.2cm}{Student generalizes code by defining variable names instead of hard-coding numbers.} & 
\parbox[t]{5.5cm}{\texttt{m\_head = 6 \# Approximated average mass}} \\
\colrule

\textbf{IV. Systems Thinking} & & \\
16. Investigating System & 
\parbox[t]{7.2cm}{Student investigates or describes the system at a macro level, focusing on overall properties. \textit{Note: A single object interacting with an external environment (gravity, friction) counts as a system.}} & 
\parbox[t]{5.5cm}{``We will solve for the average force with a point particle system."} \\

17. Relationships & 
\parbox[t]{7.2cm}{Student explains how system components interact (e.g., how friction or gravity affects motion).} & 
\parbox[t]{5.5cm}{``By setting the torque from the arm equal to the torque from gravity... we can find the mass."} \\

18. Thinking in Levels & 
\parbox[t]{7.2cm}{Student connects micro-level concepts to macro behavior. \textit{Note: Must contain a micro-level physics term and a macro-level system outcome co-located within the same text block.}} & 
\parbox[t]{5.5cm}{``The effects of drag result in our ball not going as far... our ball lands short of our desired zone."} \\

19. Communicating Info & 
\parbox[t]{7.2cm}{Student effectively presents and interprets the system’s behavior for the reader.} & 
\parbox[t]{5.5cm}{\texttt{plt.fill\_between(..., label='Valid mass range')}} \\

20. Managing Complexity & 
\parbox[t]{7.2cm}{Student defines system boundaries or makes simplifying assumptions to manage complexity.} & 
\parbox[t]{5.5cm}{``The system is the train and each of the four wheels... considered surroundings along with Earth."} \\

\end{tabular}
\end{ruledtabular}
\end{table*}

\section{Methods}

\subsection{Context and Participants} This study was conducted during the Spring 2025 semester of a calculus-based introductory mechanics course at a large Midwestern research university. The course serves as a primary requirement for engineering majors, with a total enrollment of $N=2086$ students, of which the majority are first-year students. The curriculum follows \textit{Matter and Interactions} \citep{chabay2015matter}, organizing the mechanics into three main conservation principles: Momentum, Energy, and Angular Momentum. This curricular structure is critical to the study as it emphasizes systems thinking and iterative application of fundamental laws over formula retrieval.

Computational modeling was integrated throughout the semester. Students attended weekly laboratory sections (14 total) where they utilized Python inside of Google Colab \citep{google-colab} to help visualize physical phenomena. Early laboratories focused on isolated skills (e.g., position update using average velocity), while later laboratories focused on algorithmic integration methods such as the Euler-Cromer method for solving equations. Each lab incorporated two or more of the practices in order to begin fostering students' computational thinking.

\subsection{Task: Computational Physics Essay}
The primary data source for this study, the CPE, was administered as a capstone project over the final four weeks of the semester. Students were given time at the beginning of the laboratories as well as outside of class to work on the project. Students worked in collaborative groups of two or three to write the CPE in Google Colaboratory \citep{google-colab}, a cloud-based environment for authoring Jupyter notebooks \citep{Jupyter_project}. All data was collected following approval from the institution's ethical research review board at Purdue University (Protocol Number: IRB-2020-1007).

\vspace{0.25em}
Students were instructed: 

\begin{quote}
\small\itshape
Model a real-world system of your choice. Your model must require the synthesis of at least two of the three fundamental principles (Momentum, Energy, or Angular Momentum) to fully resolve the system's dynamics.
\end{quote}

Additionally, as students modeled a wide variety of physical systems, the assignment explicitly required them to adopt standard scientific conventions. For example, students were required to explicitly define their system boundaries and uniformly treat ``up'' as the positive direction across all kinematic interactions.
To scaffold the transition from ``homework problem" to ``computational modeling," we provided students a structured Jupyter Notebook template. Unlike typical open-ended coding assignments, this template guided students through a four-stage iterative process based on the prior week's work.

\begin{enumerate}
    \item \textbf{Problem Statement}: Student defines a real-world problem and system, identifying necessary physics principles.
    \item \textbf{First Iteration/Solution}: Students construct an initial model to simulate the system.
    \item \textbf{Second Iteration/Solution}: Students must critique or improve upon their first model and implement refinement.
    \item \textbf{Conclusion}: Students synthesize their findings into a scientific argument.
\end{enumerate}

\begin{figure}[htbp]
  \centering
  \makebox[\linewidth][c]{%
    \fbox{\includegraphics[width=0.80\linewidth]{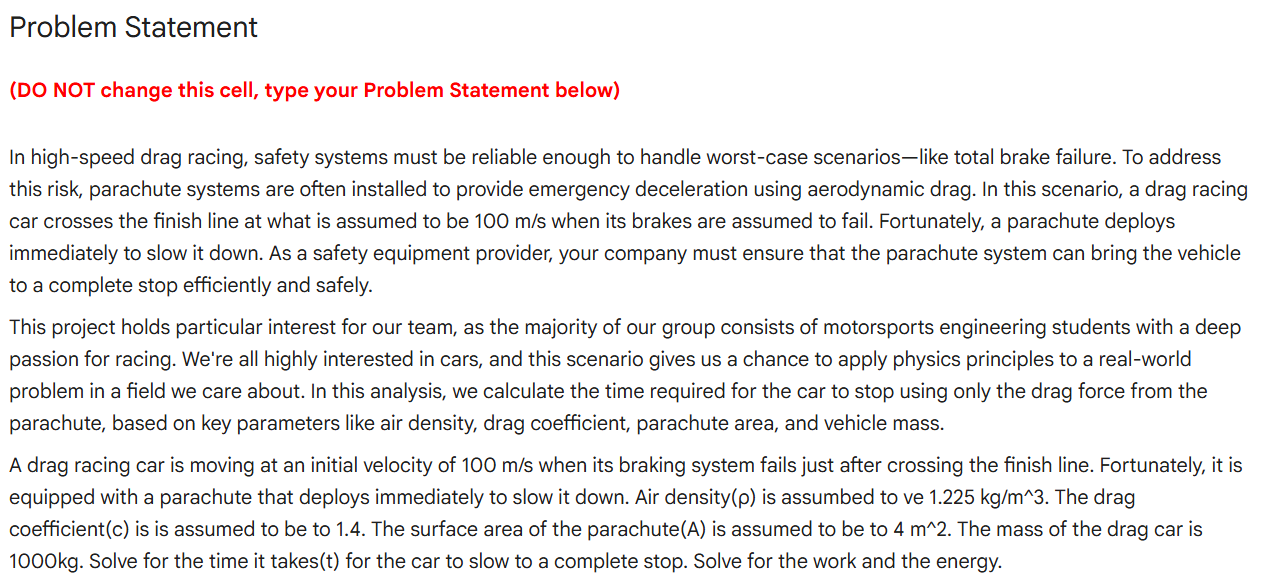}}%
  }\par\vspace{2pt}
  \makebox[\linewidth][c]{%
    \fbox{\includegraphics[width=0.80\linewidth]{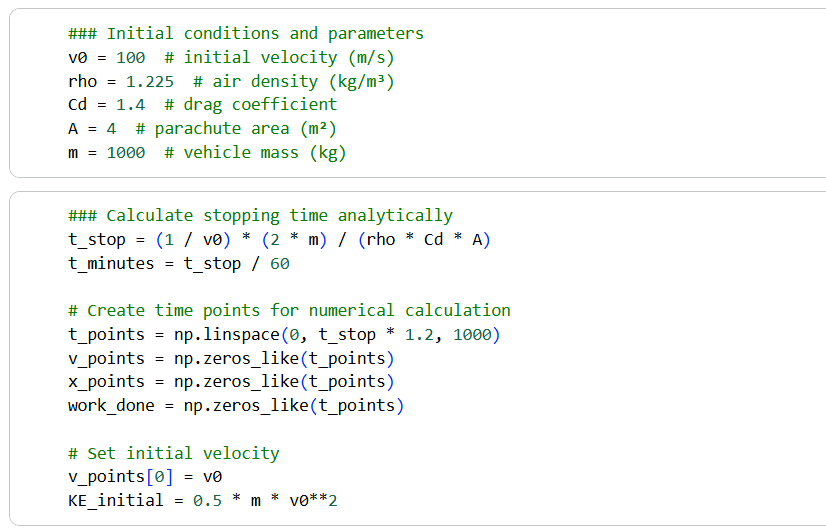}}%
  }\par\vspace{2pt}
  \caption{Example of a student's CPE Problem along with the code they used to help solve it.}
  \label{fig:exampleCPE}
\end{figure}

Common student topics included the projectile motion of objects such as rockets (energy + momentum) and the rotational motion of objects such as a figure skater (energy + angular momentum). This constraint served both to force students to define system boundaries more explicitly and also to prevent the submission of more trivial scripts. An example of a group's \textit{Problem Statement} and portion of their code can be seen in Fig.\ref{fig:exampleCPE}.

Students were evaluated for their course grade using a pedagogical rubric emphasizing scientific argumentation \citep{mcneill2011supporting}, physical accuracy, and the explicit integration of computational thinking. While this instructional rubric assessed CT across four broad domains analogous to those investigated in our study, it did not prescribe the 20 granular sub-practices measured in our subsequent research analysis. The complete student grading rubric and the structured CPE template are available upon reasonable request.
Regarding the overall volume of student output, the final submissions averaged approximately 1000 words of narrative prose, 80 lines of Python code, and 3 images per notebook. This robust combination of written text, custom programming, and data visualization supports the framing of the CPE as a comprehensive capstone project rather than a brief, superficial coding assignment.

\subsection{Data Collection}
\begin{figure}[htbp]
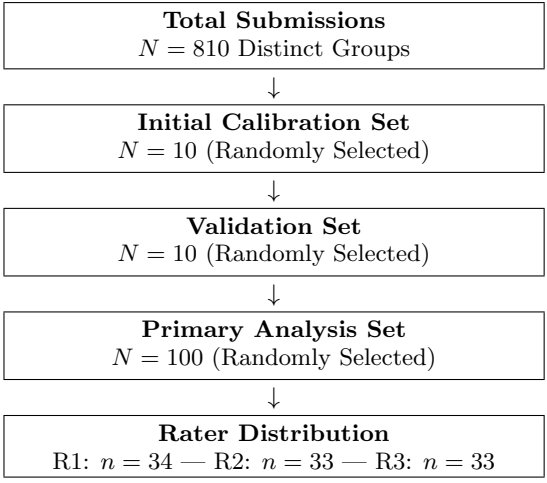

    \centering
    \renewcommand{\arraystretch}{1.25}
    \begin{tabular}{c}
        \fbox{\parbox{0.8\linewidth}{\centering \textbf{Total Submissions} \\ $N=810$ Distinct Groups}} \\
        $\downarrow$ \\
        \fbox{\parbox{0.8\linewidth}{\centering \textbf{Initial Calibration Set} \\ $N=10$ (Randomly Selected)}} \\
        $\downarrow$ \\
        \fbox{\parbox{0.8\linewidth}{\centering \textbf{Validation Set} \\ $N=10$ (Randomly Selected)}} \\
        $\downarrow$ \\
        \fbox{\parbox{0.8\linewidth}{\centering \textbf{Primary Analysis Set} \\ $N=100$ (Randomly Selected)}} \\
        $\downarrow$ \\
        \fbox{\parbox{0.8\linewidth}{\centering \textbf{Rater Distribution} \\ R1: $n=34$ | R2: $n=33$ | R3: $n=33$}} \\
    \end{tabular}
    \caption{Overview of the data-sampling and human-coding methodology.}
    \label{fig:methodology_flowchart}
\end{figure}

A total of 810 unique CPE submissions by all student groups in the class were downloaded as Jupyter Notebooks. As illustrated in Fig. \ref{fig:methodology_flowchart}, we employed stratified random sampling to select three subsets from the full dataset for our analysis:

\begin{enumerate}
    \item \textbf{Initial Calibration Set ($N=10$):} A set of 10 Jupyter notebooks was randomly selected for initial rubric development.
    
    \item \textbf{Validation Set ($N=10$):} 10 different Jupyter notebooks were randomly selected to test the inter-rater reliability (IRR) of the rubric. All three raters coded this set identically.
    
    \item \textbf{Analysis Set ($N=100$):} Following the establishment of reliability, a separate and unique set of 100 Jupyter notebooks was randomly selected for primary analysis. These Jupyter notebooks were divided into 34/33/33 and coded by three human raters.
\end{enumerate}

\subsection{Rubric Development}
To quantify the sub-practices above, we developed a binary (0/1) coding scheme based on an iterative content validation process. Initial definitions were constructed using the framework provided by \citep{weintrop2016definingcomputationalthinking}. During the calibration phase, raters used an initial set of 10 Jupyter notebooks to modify these definitions, adapting them to better fit the nature of the assessment. During this stage, raters decided to remove two categories from the framework for investigation, \textit{Programming} and \textit{Troubleshooting/Debugging}, based on the assessment.

After refining the definitions, the three expert raters independently coded a separate validation set of 10 different Jupyter notebooks across the 20 remaining sub-practices. While an inter-rater reliability (IRR) sample of $N=10$ is modest, it is consistent with qualitative coding heuristics for evaluating highly complex, multi-page artifacts. The initial independent coding yielded a raw agreement of 81\% (Fleiss' $\kappa = 0.45$) for the 200 binary data entries. Agreement on the overall holistic quality score was 73\% (Fleiss' $\kappa = 0.43$). We explicitly acknowledge that these moderate kappa values reflect the inherent subjectivity and ambiguity of evaluating open-ended computational essays. Because independent coding alone could not establish a sufficiently rigorous ground truth, these initial metrics served strictly as a diagnostic baseline. To resolve this ambiguity and establish a true gold standard, the raters subsequently met to review each sub-practice for instances of divergence. The raters discussed the evidence within the Jupyter notebooks until they reached 100\% consensus on all sub-practices for the validation set.

While these moderate kappa values reflect the inherent subjectivity of evaluating complex, open-ended computational essays, they served strictly as a diagnostic baseline. To ensure rigorous reliability, the raters subsequently met to review each sub-practice for instances of divergence. The raters discussed the evidence within the Jupyter notebooks until they reached 100\% consensus on all subpractices for the validation set. 

A key outcome of this consensus meeting was the addition of strict negative constraints to the rubric definitions. These constraints clearly distinguish between superficial computation and authentic CT. For instance, a constraint was added specifying that students who simply stated they ``imported data from GPT" would not receive credit for \textit{Collecting Data}. With these finalized constraints and a shared understanding established through consensus, the raters proceeded to independently code their own subsets of 100 different Jupyter notebooks for the primary analysis. The final definitions used can be found in Table \ref{tab:rubric_full}.

\subsection{LLM Analysis}

To determine whether generative AI could replicate the expert human evaluation, we utilized the GPT-5.4-mini API \citep{openai2025chatgpt}. Because CPEs are inherently multimodal, consisting of Python code, Markdown text, LaTeX equations, and generated data visualizations, we leveraged the model's advanced multimodal vision capabilities. 

Rather than providing the model with raw, unformatted \texttt{.ipynb} files, which can introduce excessive formatting noise, we implemented a two-stage data processing pipeline. First, the model processed the rendered visual context of each notebook by converting embedded graphs and images into detailed textual descriptions. These images were passed concurrently to the GPT-5.4-mini API using a zero-shot prompt that instructed the model to act as a physics education researcher and describe each visual in one to three sentences. 

To ensure the integrity of this transcription step, a human coder checked a random subsample of 10 notebooks, confirming the accuracy of the generated descriptions. Second, all original text, executable code, and newly generated image descriptions were compiled into a standardized CSV dataset. The LLM was then loaded with this compiled CSV file to execute the analysis strictly on the primary set of 100 notebooks. In accordance with established best practices for reducing model variance \citep{kortemeyer2023}, this entire evaluation process was repeated three independent times.

The model was configured with a temperature parameter of $0.2$ to prioritize deterministic, consistent scoring. We employed a structured prompt that utilized role-playing (instructing the model to act as an expert physics education researcher) and chain-of-thought reasoning \citep{Chen2023} for CSV analysis. Crucially, the model was provided with the exact rubric definitions, examples, strict negative constraints, and scoring criteria established during our human consensus-coding phase. An excerpt of the system prompt is provided below:

\begin{quote}
\footnotesize\ttfamily
MEGA\_SYSTEM\_PROMPT = """ You are an expert physics education researcher acting as a strict human rater. Analyze the students' Computational Physics Essay below to assess the use of Computational Thinking (CT). 

**CRITICAL RULE - CHAIN OF THOUGHT:** You MUST extract and write out the specific quote or code snippet that justifies your score in the "Reasoning" field BEFORE outputting any numerical scores. 

--- SCORING PRESENCE (0 or 1) --- 

**0 (Absent):** The practice is NOT observed, violates a specific rubric calibration constraint, or requires the rater to guess the student's intent. If a practice is only vaguely implied, or if you are in doubt score 0. 

**1 (Present):** The rater can easily point to an explicit use of the practice. 

--- SCORING QUALITY (0-2) --- 
**Apply this score to BOTH the individual Domain "Quality" AND the "Overall\_Quality".** 

**0 (None/Broken):** Practices are absent, or the notebook/code is broken, incomplete, or lacks a clear purpose. 

**1 (Isolated):** Working code or math or physics is present. There may be some markdown text, but it is superficial or generic. When in doubt between a 1 and a 2, default to a strict 1. 

**2 (Integrated):** Code/math works AND is accompanied by text that directly connects the specific computational outputs back to the physical constraints of the problem to draw a conclusion. Do NOT award a 2 for generic summaries or basic observations..... """
\end{quote}


\section{Results \& Discussion} 

\subsection{Human Baseline}

To ensure the validity of our assessment, we first evaluated the reliability of the human coding process. During the initial calibration phase, the three raters independently coded a validation set of 10 Jupyter notebooks, achieving an average notebook score of $15.4 \pm 1.2$ out of 20 (mean and standard deviation). As for overall quality, the average notebook score was $1.4 \pm 0.1$ out of 2.0. Following this calibration, we examined the scores across the three independently coded subsets of the main dataset to check for rater drift, a phenomenon where raters diverge when grading independently \citep{hoskens2001realtime}.

Among the 100 Jupyter notebooks in the analysis set, the population distribution yielded an average CT coverage score of $16.51 \pm 2.54$ out of 20, and an average CT Quality score of $1.62 \pm 0.52$. To verify that no rater drift occurred during this phase, we compared the individual means across the three raters' respective subsets ($n \approx 33$). The subset means clustered well, with a standard deviation of just $0.02$ and $0.10$ for CT coverage and Quality between the three raters, respectively. This consistency provides a secondary validation of our methodology, indicating both that the random sampling distributed the population uniformly and that the raters maintained strict adherence to the rubric without shifting excessively.

Across the full analysis set, students demonstrated high levels of engagement with the prompt. As illustrated by the human consensus distribution in Fig. \ref{fig:histogramofCT}, the scores are heavily right-skewed, with a majority of groups demonstrating the use of over 75\% of the desired sub-practices. 

To provide granular insight into the specific actions students took within their Jupyter notebooks, Table \ref{tab:subpractice_percentages} displays the proportion of student groups that successfully demonstrated each of the 20 sub-practices. The most observed practice was \textit{Investigating a Complex System as a Whole} (99\%), confirming the ``real-world problem'' constraint successfully forced students to adopt systems-level perspectives. Conversely, \textit{Developing Modular Computational Solutions} was the least observed practice (50\%), highlighting a gap in students' deeper software engineering skills.

\subsection{LLM Analysis of CT Sub-Practices}

We evaluated the LLM's capacity to detect the 20 binary CT sub-practices. Looking at the LLM's independent performance across the three evaluation runs, the model demonstrated high internal consistency. As shown in Table \ref{tab:subpractice_percentages} ($LLM \pm SD$), the standard deviation across the three runs was remarkably low for the majority of categories. The only notable internal variance occurred in \textit{Manipulating Data} and \textit{Collecting Data}, likely due to the LLM experiencing semantic confusion between students transforming data versus gathering it, or the constraint of \textit{Citations of LLMs (e.g. ChatGPT) do NOT count}. 

When comparing the automated outputs directly to the human ground truth, the LLM achieved 84\% agreement overall. The model exhibited exceptionally high alignment on structural and code-based practices, such as \textit{Constructing Computational Models} (96\% agreement) and \textit{Creating Computational Abstractions} (98\% agreement). 

\begin{figure}[htbp]
    \centering
    \includegraphics[width=.95\linewidth]{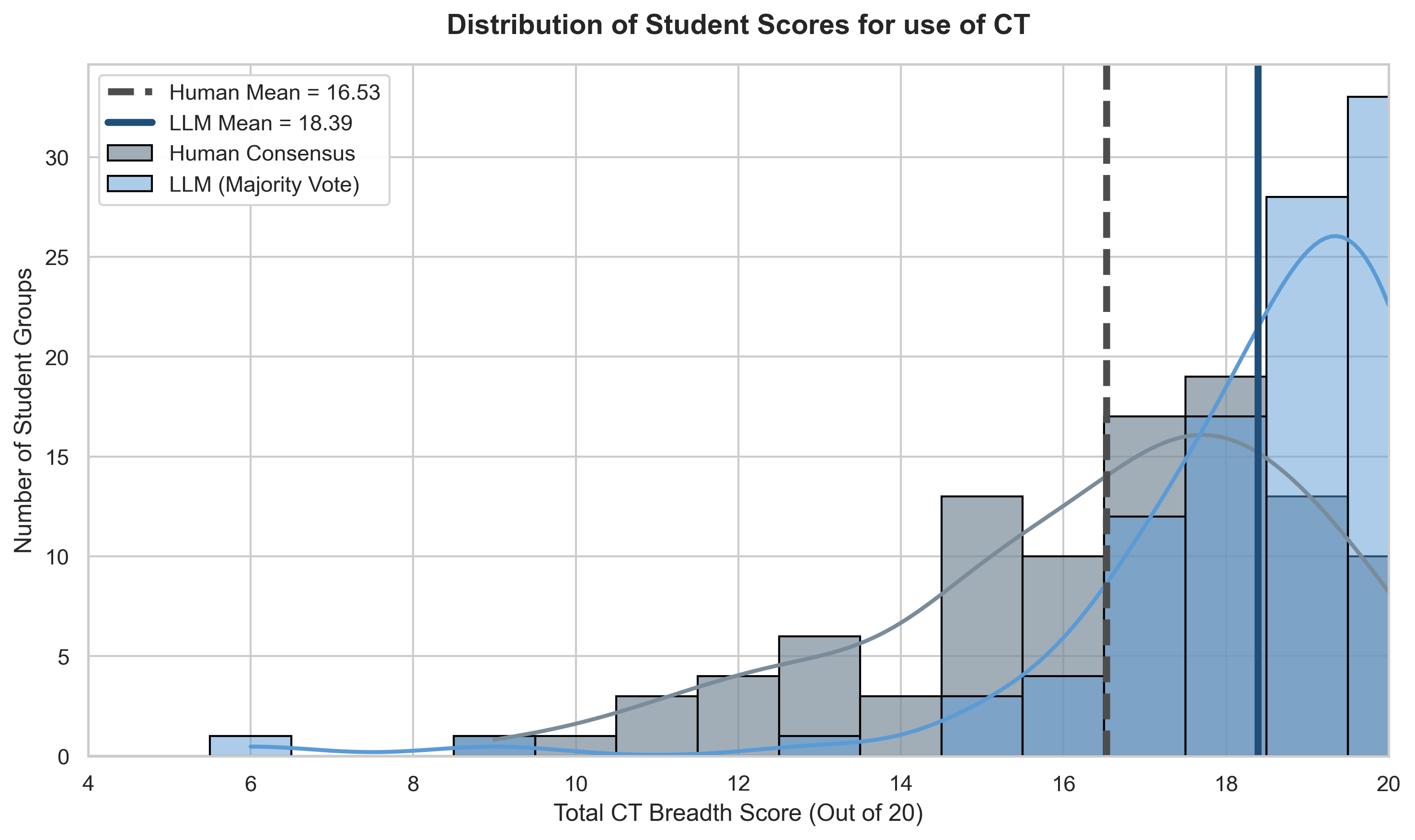}
    \caption{Distribution of students' overall Computational Thinking breadth scores (maximum of 20 points) for both human and LLM.}
    \label{fig:histogramofCT}
\end{figure}

However, assessing the true reliability of this agreement requires careful statistical framing. When evaluating inter-rater reliability, Cohen's $\kappa$ can become artificially deflated when the data is heavily skewed toward one category. Because our student population performed very well (as shown by the right-skewed distribution in Fig. \ref{fig:histogramofCT}), the vast majority of scores were `1' rather than `0'. This uneven distribution triggers a well-documented statistical phenomenon known as the ``kappa paradox'' \citep{feinstein1990high, sim2005kappa}, where high exact agreement mathematically yields a disproportionately low $\kappa$ value. To account for this paradox and prevent artificially low reliability scores, we utilized the Prevalence-Adjusted Bias-Adjusted Kappa (PABAK) \citep{byrt1993bias}. When evaluated through the lens of PABAK, the multimodal LLM demonstrated substantial agreement (PABAK $\ge 0.6$) with human consensus across 15 of the 20 sub-practices and an average PABAK $= 0.69$.

To better understand the boundary conditions of the automated evaluation, we conducted a qualitative check of the LLM's generated reasoning, focusing on both its successes and its failures. The LLM performed well at identifying objective, structural practices such as detecting \textit{Understanding Relationships within a System} (99\% exact agreement) when students explicitly linked variables. In Notebook 110492, the LLM successfully identified the physical sensemaking:

\begin{quote}
\footnotesize\itshape
Quote evidence: ``The air resistance force increases as the velocity squared, which eventually balances gravity so the rocket reaches terminal velocity.'' \\
LLM Rationale: ``The student explicitly connects the interaction of air resistance and velocity to the overarching macro-level behavior of the system (terminal velocity).''
\end{quote}

In these highly structured contexts, the model successfully reproduces rubric-based observations. However, evaluating subjective epistemic agency remains a challenge. To investigate where the LLM diverged most from expert evaluation, we analyzed the sub-practice with the lowest inter-rater reliability: \textit{Choosing Effective Computational Tools} ($\kappa = 0.02$, PABAK = 0.24). 

The LLM exhibited a severe positive bias here, awarding this practice to 93\% of groups compared to the human baseline of 63\%. A review of the model's generated reasoning reveals a persistent semantic vulnerability. The expert human rubric defines this practice as the student explicitly \textit{justifying} their selection of libraries or physics approaches. However, the LLM frequently misclassified the mere descriptive mention of a tool as a justified choice. For example, in Notebook 304029, the human raters scored the practice as absent (0) because the student provided no underlying rationale. The LLM scored it as present (1), formulating the following rationale:

\begin{quote}
\footnotesize\itshape
Quote evidence: ``numpy for mathematical calculations'' and ``matplotlib.pyplot for data visualization''. \\
LLM Rationale: ``These show planning, tool choice, iterative comparison, modularization, and abstraction with named parameters.''
\end{quote}

This discrepancy highlights a boundary condition in which LLMs' grading defaults to description over evaluation. The model executed \textit{structural recognition} well, identifying the exact presence of the tools (NumPy and Matplotlib) and correctly mapping them to their standard Python functions. However, it completely failed to measure the student's \textit{epistemic agency}. It confused the existence of syntax with scientific justification, failing to capture the ``why'' that human raters demand when assessing authentic computational thinking. Though strict negative constraints successfully guided the LLM in objective categories, this structural blindness to deeper scientific justification remains a significant hurdle for grading tools.

\begin{table*}[htbp]
\caption{Comparison of Human and LLM Scoring for CT Sub-practices ($N=100$). The table presents the proportion of student groups engaging in each practice (0-1 scale) as coded by humans and the LLM (Mean $\pm$ SD across 3 runs), alongside reliability metrics (Agreement \%, and PABAK).}

\label{tab:subpractice_percentages}
\renewcommand{\arraystretch}{1.6} 
\scriptsize
\begin{ruledtabular}
\begin{tabular}{l c c c c c} 
\textbf{Computational Thinking Sub-practice} & \textbf{Human Mean} & \textbf{LLM $\pm$ SD} & \textbf{Agree (\%)}  & \textbf{PABAK} \\ 
\colrule

\textbf{I. Data Practices} & & & & & \\ 
Creating Data & 0.86 & $0.97 \pm 0.04$ & 87  & 0.74 \\
Manipulating Data & 0.83 & $0.84 \pm 0.16$ & 79  & 0.58 \\
Collecting Data & 0.79 & $0.60 \pm 0.20$ & 77  & 0.54 \\
Visualizing Data & 0.79 & $0.83 \pm 0.02$ & 94 & 0.88 \\
Analyzing Data & 0.75 & $0.97 \pm 0.03$ & 72  & 0.44 \\ 
\colrule

\textbf{II. Modeling \& Simulation Practices} & & & & & \\ 
Constructing Computational Models & 0.95 & $0.99 \pm 0.00$ & 96  & 0.92 \\
Using Computational Models to Find and Test Solutions & 0.91 & $0.98 \pm 0.01$ & 93 & 0.86 \\
Assessing Computational Models & 0.91 & $0.97 \pm 0.02$ & 90 &  0.80 \\
Designing Computational Models & 0.87 & $0.96 \pm 0.08$ & 83  & 0.66 \\
Using Computational Models to Understand a Concept & 0.85 & $0.98 \pm 0.01$ & 87 & 0.74 \\ 
\colrule

\textbf{III. Computational Problem-Solving Practices} & & & & & \\ 
Creating Computational Abstractions & 0.97 & $0.99 \pm 0.01$ & 98  & 0.96 \\
Assessing Different Approaches/Solutions to a Problem & 0.86 & $0.93 \pm 0.07$ & 81 & 0.62 \\
Preparing Problems for Computational Solutions & 0.85 & $0.95 \pm 0.09$ & 80  & 0.60 \\
Choosing Effective Computational Tools & 0.63 & $0.93 \pm 0.06$ & 62  & 0.24 \\
Developing Modular Computational Solutions & 0.50 & $0.55 \pm 0.08$ & 83 & 0.66 \\ 
\colrule

\textbf{IV. Systems Thinking Practices} & & & & & \\ 
Investigating a Complex System as a Whole & 0.99 & $0.98 \pm 0.02$ & 80  & 0.60 \\
Understanding Relationships within a System & 0.89 & $1.00 \pm 0.01$ & 99 & 0.98 \\
Defining Systems and Managing Complexity & 0.87 & $1.00 \pm 0.01$ & 86  & 0.72 \\
Communicating Information about a System & 0.79 & $1.00 \pm 0.00$ & 89 & 0.78 \\
Thinking in Levels & 0.67 & $0.97 \pm 0.10$ & 70  & 0.40 \\ 
\colrule

\textbf{Overall Metrics} & & & & & \\
Overall CT Quantity (0--20) & 16.51 & $18.39 \pm 0.69$ & 84 &  0.69 \\
\end{tabular}
\end{ruledtabular}
\end{table*}

\subsection{LLM Analysis of Holistic Quality Scores}

Beyond the binary presence of sub-practices, our second research question evaluated the LLM's capacity to assess the holistic, subjective quality of the student essays on a 0-to-2 scale. Table \ref{tab:quality_metrics} presents a comparison of the human consensus quality scores and the LLM's average scores across the four overarching domains and the overall assessment. 

As shown in the table, the LLM's assessment closely tracked human intuition in the highly structured domains, achieving near-perfect average alignment in \textit{Data Practices} (1.61 vs. 1.65) and \textit{Modeling Practices} (1.72 vs. 1.71). The model exhibited a slight positive scoring bias in \textit{Problem Solving} (1.60 vs. 1.53) and a larger positive bias in \textit{Systems Thinking} (1.84 vs. 1.63). This larger bias likely occurs because the model conflates the occurrence of vocabulary with actual conceptual synthesis. While human raters demand explicit connections between system components, the LLM frequently over-credits students who simply locate relevant system terminology within the same text block. Ultimately, this resulted in an \textit{Overall CT Quality} average of 1.77 for the LLM, compared to the human baseline of 1.62.

To measure inter-rater reliability for this ordinal data, we calculated Krippendorff's $\alpha$ \citep{hayes2007answering} (Table \ref{tab:quality_metrics}). The overall $\alpha = 0.39$ indicates weak absolute agreement. However, because students performed exceptionally well, scores are heavily concentrated in the `1' and `2' categories. This lack of variance mathematically shrinks $\alpha$, causing it to appear artificially low despite high agreement. 

To better quantify grading discrepancies, we evaluated Exact Agreement and Mean Absolute Error (MAE). On a 0-2 scale, an MAE below 0.5 demonstrates acceptable reliability, indicating the model deviates by less than half a point on average \citep{Baccianella2009}. As shown in Table \ref{tab:quality_metrics}, the LLM achieved 71\% exact agreement overall with an MAE of just 0.29. Even in highly subjective domains like \textit{Systems Thinking} and \textit{Problem Solving}, the MAE remained $\le 0.37$. Thus, when the LLM diverges from human consensus, it rarely misses by more than a fraction of a point, ensuring the automated scores remain instructionally valid.

\begin{table*}[htbp]
\caption{Inter-rater reliability metrics for holistic Computational Thinking Quality scores (0-to-2 scale) across the four core domains and the overall assessment ($N=100$).}
\label{tab:quality_metrics}
\renewcommand{\arraystretch}{1.4} 
\begin{ruledtabular}
\begin{tabular}{l c c c c c}
\textbf{CT Quality Domain} & \textbf{Human ($M \pm SE$)} & \textbf{LLM ($M \pm SE$)} & \textbf{Exact (\%)} & \textbf{MAE}  \\
\colrule
Data Practices & 1.65 $\pm$ 0.06 & 1.61 $\pm$ 0.05 & 65.0\% & 0.36  \\
Modeling Practices & 1.71 $\pm$ 0.05 & 1.72 $\pm$ 0.05 & 71.0\% & 0.29  \\
Problem Solving & 1.53 $\pm$ 0.05 & 1.60 $\pm$ 0.05 & 63.0\% & 0.37  \\
Systems Thinking & 1.63 $\pm$ 0.05 & 1.84 $\pm$ 0.04 & 67.0\% & 0.33  \\
\colrule
\textbf{Overall CT Quality} & \textbf{1.62 $\pm$ 0.05} & \textbf{1.77 $\pm$ 0.04} & \textbf{71.0\%} & \textbf{0.29}  \\
\end{tabular}
\end{ruledtabular}
\end{table*}

Across \textit{Data Practices, Modeling and Simulation and Computational Problem Solving} domains, the standard error of the mean for the LLM's quality scores is similar to the human consensus, as shown in Table \ref{tab:quality_metrics}, with higher values for \textit{Systems Thinking and Overall CT}. Unlike the binary checklist, the holistic metric requires the rater to evaluate the depth of physical sensemaking and the integration of a narrative argument. The larger standard error suggests that the LLM is highly sensitive to semantic noise in these essays, where slight variations in how a student phrased a paragraph can cause the model's assigned score to fluctuate much more than a human expert's reasoning. 

This variance highlights a critical boundary condition: while grading tools excel at identifying explicit code structures, they struggle with the stability required for subjective epistemic evaluation. Yet, because the MAE remains extremely low, the aggregated scores remain instructionally valid. While this pipeline is not currently suitable for providing automated, unreviewed formative feedback directly to students—as it may mischaracterize their epistemic justifications—it serves as a highly effective tool for programmatic assessment. For instructors managing large-enrollment environments, it provides rapid, baseline metrics on class-wide CT integration and can successfully triage superficial notebooks for targeted human review.

\section{Conclusions \& Future Work}

This study demonstrated the viability of using a multimodal Large Language Model to automate the rubric-based evaluation of open-ended, yet scaffolded, Computational Physics Essays (CPEs). Addressing our first research question (RQ1), the constrained LLM pipeline demonstrated a strong capacity for reproducing rubric-based observations. While absolute Cohen's $\kappa$ values were suppressed due to the heavily skewed data distribution, the model achieved a robust 84\% exact agreement with human raters. It performed exceptionally well on explicit, structural tasks like identifying iterative loops and defining system boundaries. However, our qualitative audit revealed that the LLM exhibits semantic vulnerabilities in highly subjective categories like \textit{Choosing Effective Computational Tools}, frequently mistaking the mere presence of a coding tool for an authentic epistemic justification.

For RQ2, evaluating the holistic quality of student work proved more challenging. As is typical for subjective epistemic tasks, exact agreement was lower (71\%), and the model exhibited a slight positive scoring bias ($\alpha = 0.39$). However, its extremely low Mean Absolute Error ($\text{MAE} = 0.29$) ensures that the automated scores rarely deviate from human intuition by more than a fraction of a point. 

Ultimately, the strongest contribution of this study is demonstrating that AI can reproduce many rubric-based observations in multimodal computational essays with useful—though imperfect—agreement. While we explicitly caution against utilizing this current architecture to provide automated, unreviewed formative feedback directly to students, it serves as a highly effective tool for programmatic assessment. By providing instructors with rapid, baseline metrics on class-wide CT integration, this methodology significantly mitigates the traditional grading bottleneck, enhancing the feasibility of complex, project-based assessments in large-enrollment physics courses.

Having validated this baseline against human consensus, our immediate future work involves deploying the pipeline across the entire course population ($N>2000$). While this introduces operational constraints regarding data processing, the reduction in grading time—from dozens of hours of expert coding to mere minutes—makes programmatic, class-wide assessment scalable. We explicitly caution against utilizing this current architecture to return automated, unreviewed grades directly to students, given the model's vulnerability in parsing subjective "sensemaking" contexts. Furthermore, as we expand to diverse institutional contexts, we must actively investigate potential algorithmic bias. Because the LLM was calibrated on a localized dataset, future research must ensure the model evaluates students strictly on their physical reasoning, rather than inadvertently penalizing non-traditional demographics who may articulate their understanding using different vocabulary or linguistic markers. Ensuring this pipeline's fairness across diverse populations will be critical for equitable assessment.

\begin{acknowledgments}
This work is supported in part by the U.S. National Science Foundation under Grant NSF-2300645. Any opinions, findings, and conclusions or recommendations expressed in this material are those of the authors and do not necessarily reflect the views of the National Science Foundation. The datasets analyzed during the current study are not publicly available due to student privacy restrictions, but are available from the corresponding author on reasonable request. The authors declare no competing financial interests.
\end{acknowledgments}

\clearpage
\bibliography{bibby}
\end{document}